\documentclass[12pt, prd, aps, preprintnumbers, nofootinbib]{revtex4}

\usepackage{epsfig}
\usepackage{amsmath}
\usepackage{amssymb}

\newcommand{\be}{\begin{equation}}
\newcommand{\ee}{\end{equation}}
\newcommand{\bee}{\begin{equation*}}
\newcommand{\eee}{\end{equation*}}
\newcommand{\bea}{\begin{eqnarray}}
\newcommand{\eea}{\end{eqnarray}}
\newcommand{\bean}{\begin{eqnarray*}}
\newcommand{\eean}{\end{eqnarray*}}

\newcommand{\lp}{\left(}
\newcommand{\rp}{\right)}

\begin{document}

\preprint{UAB-FT-666, MCTP-09-10}

\title{Strong electroweak phase transitions without collider traces}

\author{A.~Ashoorioon}
\email{amjad@umich.edu}
\affiliation{Michigan Center for Theoretical Physics,
University of Michigan, Ann Arbor, Michigan 48109-1040, USA}

\author{T.~Konstandin}
\email{konstand@ifae.es}
\affiliation{Institut de F\'isica d'Altes Energies,
Universitat Aut\`onoma de Barcelona, 08193 Bellaterra, Barcelona, Spain }

\date{\today}

\begin{abstract}
We discuss the question if the upcoming generation of collider and
low-energy experiments can successfully probe the nature of the
electroweak phase transition. In particular, we are interested in phase
transitions strong enough for electroweak baryogenesis or even for
a production of gravitational radiation observable by the Big Bang
Observer.

As an explicit example, we present an analysis in a singlet extension
of the Standard Model. We focus on the region in parameter space where
the model develops no significant deviation in its low energy
phenomenology from the Standard Model. Nevertheless, this class of
models can develop a very strong phase transition.
\end{abstract}

\maketitle

\section{Introduction}

Main objective of the upcoming generation of collider experiments,
first and foremost the LHC, is to unravel the nature of electroweak
symmetry breaking. The most simplistic model that up to now fits all
experimental data is hereby the Standard Model (SM) of particle
physics that incorporates the most simple (and perturbatively
controllable) scenario: A Higgs sector with a single $SU(2)_L$ doublet
that breaks the corresponding gauge invariance by means of a vacuum
expectation value (VEV).

However, the 'discovery' of this simple scenario would lead to some
disappointments: First, the hierarchy problem of the SM suggests that
the actual symmetry breaking mechanisms is more complicated and might
incorporate supersymmetry, composite Higgs models, Little Higgs
models, extra-dimensions, and so on (for a recent review on
alternative approaches to electroweak symmetry breaking see
ref.~\cite{Grojean:2007zz}). Secondly, a SM Higgs sector lacks a
strong electroweak phase transition that is required e.g. for
electroweak baryogenesis~\cite{Farrar:1993hn}.

In the present work, we discuss the question if upcoming collider and
low-energy energy probes compatible with the SM necessary imply the
occurrence of a cross-over rather than a strong first-order phase
transition. To be specific, we discuss a singlet extension of the SM
and show that an electroweak phase transition strong enough for
electroweak baryogenesis or even sizable gravitational wave production
can in certain models be reconciled with SM collider phenomenology. In
particular, we provide a concrete numerical example for this case.

The paper is organized as follows: In sec.~\ref{sec_model} we present
the model and set up notation. We also introduce a parametrization of
the scalar potential that is more appropriate for the discussion of
the phase transition. In sec.~\ref{sec_LEprobes} we discuss possible
collider signals and the constraints on the parameter space to avoid a
distinction from the SM. We also comment on the possibility of the
singlet being a viable dark matter candidate. In secs.~\ref{sec_PT}
and \ref{sec_PTnum} we discuss the electroweak phase transition in the
model and give an explicit numerical example. We finally conclude in
sec.~\ref{sec_concl}.

\section{The Model \label{sec_model}}

The model we study in the present work is the minimal singlet
extension of the Standard Model as discussed in
refs.~\cite{Espinosa:1993bs, Choi:1993cv, Datta:1997fx,
O'Connell:2006wi, BahatTreidel:2006kx, Barger:2007im, Profumo:2007wc}
and its recent revival in form of hidden sector
theories~\cite{Schabinger:2005ei, Espinosa:2007qk,
Espinosa:2008kw}. The associated tree-level scalar potential reads
\footnote{We mostly adhere to the notation of ref.~\cite{Profumo:2007wc}
that partially parallels our analysis.  }
\be
V = V_{SM} + V_{HS} + V_S
\ee
with
\bea
V_{SM} &=& - \mu^2 \, H^\dagger H + \lambda_0 (H^\dagger H)^2, \\
V_{HS} &=& \frac{a_1}{2} \, H^\dagger H \, S
   + \frac{a_2}{2} \, H^\dagger H \, S^2, \\
V_S &=& b_1 S + \frac{b_2}{2} S^2 + \frac{b_3}{3} S^3 + \frac{b_4}{4} S^4,
\eea
where $H$ denotes the SM $SU(2)_L$ scalar Higgs doublet and $S$
denotes the additional scalar singlet. We could shift the singlet
field in order to remove the linear contribution from the potential,
but it will prove useful to keep it in the following for the
regularization of the potential.

In order to discuss the symmetry breaking pattern, we rewrite the
fields in terms of vacuum expectation values (VEVs) and fluctuations,
using for the neutral Higgs component $H=(v+h)/\sqrt{2}$ and for the
singlet field $S=x+s$, leading to
\be
\label{pot_old}
V = - \frac{\mu^2}{2} \, v^2 + \frac{\lambda_0}{4} v^4
+ \frac{a_1}{4} \, v^2 \, x + \frac{a_2}{4} v^2 \, x^2
+ b_1 x + \frac{b_2}{2} x^2 + \frac{b_3}{3} x^3 + \frac{b_4}{3} x^4.
\ee
In the case $a_1=b_1=b_3=0$, the potential has a $\mathbb{Z}_2$ symmetry and
the singlet is stable as long this symmetry is not spontaneously
broken. However, these conditions are not necessary as we will see in a
later section.

The aim of the present paper is to focus on regions in parameter space
in which the model develops a strong first-order electroweak phase
transition. It turns out that the tree-level potential is in this
context better characterized by the following eight conditions that
can by traded for the eight parameters in eq.~(\ref{pot_old}), namely
the definition of the two VEVs of the broken phase $\bar v$ and $\bar
x$
\be
\label{constr1}
0 = \left. \frac{\partial V}{\partial v} \right|_{v=\bar v,\, x=\bar x}, \quad
0 = \left. \frac{\partial V}{\partial x} \right|_{v=\bar v,\, x=\bar x},
\ee
the tree-level Higgs and singlet masses and mixing
\be
\label{constr2}
\bar \mu_h^2 = \left. \frac{\partial^2 V}{\partial v^2} \right|_{v=\bar v,\, x=\bar x}, \quad
\bar \mu_s^2 = \left. \frac{\partial^2 V}{\partial x^2} \right|_{v=\bar v,\, x=\bar x}, \quad
\bar \mu_{hs}^2 = \left. \frac{\partial^2 V}{\partial v \partial x} \right|_{v=\bar v,\, x=\bar x},
\ee
the singlet VEV of the symmetric phase (which by convention we choose
to vanish) and the masses of the Higgs and singlet in the symmetric
phase
\be
\label{constr3}
0 = \left. \frac{\partial V}{\partial x} \right|_{v=0,\, x=0}, \quad
\mu_h^2 = \left. \frac{\partial^2 V}{\partial v^2} \right|_{v=0,\, x=0}, \quad
\mu_s^2 = \left. \frac{\partial^2 V}{\partial x^2} \right|_{v=0,\, x=0}.
\ee
Notice that the derivative $\partial V/ \partial v |_{v=0, \, x=0}$
and $\partial^2 V/ \partial v \partial x |_{v=0, \, x=0}$
automatically vanish such that by these eight parameters the potential
is up to second derivatives completely constrained in the broken and
symmetric phase.

To replace the eight parameters in eq.~(\ref{pot_old}) by these
constraints is always possible, as long as the singlet VEV $\bar x$
does not vanish. In this limit, some of the original parameters could
approach infinity as can been seen by their explicit form as given in
the appendix. In the examples we make sure that the parameters are
small enough to treat the model perturbatively. We also check the
correlation between the parameters we use here and the parameters in
eq.~(\ref{pot_old}) to ensure that no tuning is embodied by our choice
of parameters in the specific numerical examples.

The advantages of this kind of parametrization are apparent: First,
with the VEVs and the masses and mixing, the potential is described by
physical quantities that will enter in the phenomenological
discussion. Second, the second derivatives of the potential in the
symmetric phase, $\mu_h^2$ and $\mu_s^2$ will lead to two local minima
in the potential if chosen positive. Hence, the symmetric phase is
even at zero temperature separated from the broken phase by a
potential barrier. This is essential for a very strong phase
transition.

Even though this model can develop a barrier between the two local
minima already at tree-level, one-loop contributions to the potential
(and the free energy) are essential to discuss the temperature
dependence and we hence include the Coleman-Weinberg
contributions~\cite{Coleman:1973jx} for consistency, that read
\bea
V_1 &=& \sum_k \, n_k \, G[m^2_k] + V_\textrm{counterterms}, \\
G(y) &=& \frac{y^2}{64\pi^2} \left[ \ln \left(\frac{y}{Q^2}\right)
- \frac32\right].
\eea
Usually the counterterms are arranged as to enforce the conditions in
eq.~(\ref{constr1}). For example, in supersymmetric theories, the soft
mass terms can be shifted accordingly which subsequently results in
the modification of the physical Higgs mass. However, the present
model is not constrained by supersymmetry and we can enforce all
constraints eqs.~(\ref{constr1})-(\ref{constr3}) in our
renormalization prescription (see appendix). This has several
advantages: First, the physical masses of the Higgs and singlet
coincide on tree-level and one-loop level (also the Goldstone bosons
remain massless which leads to a technical difficulty that is
discussed in ref.~\cite{Delaunay:2007wb}). Secondly, the relevant
features of the potential to discuss the phase transition are
unchanged by the one-loop contributions. Last, in case of a heavy
singlet, the influence of the singlet on the $v$-dependence of the
potential close to the broken minimum automatically ceases and the
singlet 'decouples'.

At finite temperature, the effective potential receives in addition
the following one-loop contributions
\be
V_1^{T\not=0} = \frac{T^4}{2 \pi^2} \sum_k n_k J_{B/F}
[(m_k^2 + \Pi_k)/T^2],
\ee
with
\be
J_{B/F} (y) = \int_0^\infty dx \, x^2 \, \ln
\left[ 1 \mp \exp \left(-\sqrt{x^2 + y} \right) \right],
\ee
and the thermal self-energies $\Pi_k$. Notice that expanding the
expression in the thermal-energies gives rise to the more common
one-loop expression without the self-energies and the Daisy-diagrams
as detailed in ref.~\cite{Delaunay:2007wb}. The thermal self-energies
are given in the appendix.

\section{Low Energy Probes of the Model \label{sec_LEprobes}}

In this section we discuss several low-energy probes of the
model. First, consider the electroweak precision observables (EWPO)
that also have been studied in the model at hand in
ref.~\cite{Profumo:2007wc, Barger:2007im}. In leading order, most
deviations from the SM values result from the mixing between the Higgs
and the singlet. With above given mass terms and the mass eigenstates
\be
h_1 = \sin \theta \, s + \cos \theta \, h, \quad
h_2 = \cos \theta \, s - \sin \theta \, h,
\ee
the mixing parameter is given by
\be
\tan \theta = \frac{y}{1 + \sqrt{1 + y^2}}, \quad
y = \frac{\bar \mu^2_{hs}}{\bar \mu_h^2 - \bar \mu_s^2}.
\ee
Then, the deviation of the $T$ parameter from its SM value is in
leading order e.g. given by
\bea
T- T_{SM} &=& \lp \frac{3}{16 \pi \sin \theta_W^2} \rp
\left\{
\frac{m_Z^2}{ m_W^2} \lp \frac{m_h^2}{m_h^2 - m_Z^2} \rp
\log \frac{m_h^2}{m_Z^2}
-                     \lp \frac{m_h^2}{m_h^2 - m_W^2} \rp
\log \frac{m_h^2}{m_W^2}  \right. \nonumber \\
&& - \cos^2 \theta \left[  \frac{m_Z^2}{ m_W^2}
\lp \frac{m_1^2}{m_1^2 - m_Z^2} \rp
\log \frac{m_1^2}{m_Z^2}
-                     \lp \frac{m_1^2}{m_1^2 - m_W^2} \rp
\log \frac{m_1^2}{m_W^2}  \right] \nonumber \\
&& - \left. \sin^2 \theta \left[  \frac{m_Z^2}{ m_W^2}
\lp \frac{m_2^2}{m_2^2 - m_Z^2} \rp
\log \frac{m_2^2}{m_Z^2}
-                     \lp \frac{m_2^2}{m_2^2 - m_W^2} \rp
\log \frac{m_2^2}{m_W^2}  \right]
\right\}.
\eea
Hence, if mixing between the Higgs and the singlet is suppressed, the
singlet extension cannot be distinguished from the SM with Higgs mass
$m_h=m_1$. So without further ado we use the freedom in the potential
to set the mixing to zero, $\bar \mu^2_{hs} = 0$.

\begin{figure}[ht]
\includegraphics[width=0.9\textwidth, clip ]{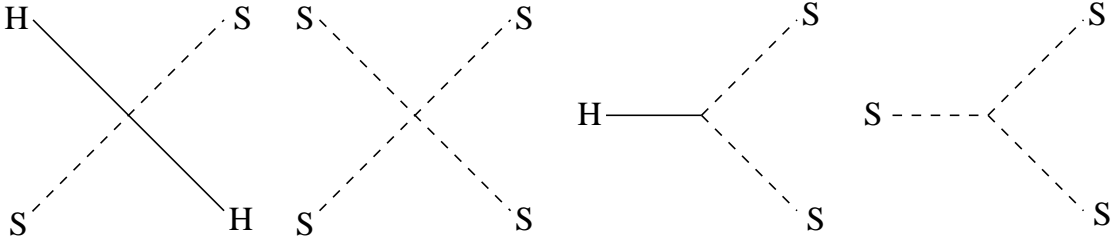}
\caption{\label{fig:feyn}
The additional Feynman rules involving the singlets under the
assumption of vanishing Higgs-singlet mixing, $\bar \mu^2_{hs} = 0$.}
\end{figure}

Next, we consider the detection of the additional degree of freedom in
collider experiments. In order to do so, consider the Feynman rules
arising from the tree-level potential. In particular, we assume that
there is no mixing between the Higgs and the singlet, $\bar \mu^2_{hs}
= 0$, to reduce the constraints on the EWPO as explained before.

For the Higgs self-interactions one finds at tree-level
\be
\left. \frac{\partial^3 V}{\partial v^3} \right|_{v=\bar v,\, x=\bar x}
= \frac{3 \bar \mu_h^2}{ \bar v}, \quad
\left. \frac{\partial^4 V}{\partial v^4} \right|_{v=\bar v,\, x=\bar x}
= \frac{3 \bar \mu_h^2}{ \bar v^2},
\ee
and hence exactly the SM relations. There will be a small modification
due to the additional one-loop contribution of the singlet, but this
effect is too small to be measured in the near future. Notice that
this is somewhat in contradiction with the claim in
ref.~\cite{Noble:2007kk} that a strong phase transition necessarily
leaves traces in the Higgs self-couplings (see also
ref.~\cite{Espinosa:2008kw} for the corresponding discussion with many
strongly coupled singlets and refs.~\cite{Kanemura:2004ch,
Aoki:2008av} for an analysis in a two Higgs doublet model).

Compared to the SM there are new interaction terms with the singlet
which for vanishing mixing, $\bar \mu_{hs}=0$, read
\bea
\left. \frac{\partial^3 V}{\partial x^2 \partial v}
\right|_{v=\bar v,\, x=\bar x}
&=& \frac{(2 \mu^2_h + \bar \mu_h^2)\bar v}{ \bar x^2}, \\
\left. \frac{\partial^3 V}{\partial x^3}
\right|_{v=\bar v,\, x=\bar x}
&=& \frac{ (2 \mu_s^2 + 4 \bar \mu_s^2) \bar x^2
- (4 \mu_h^2 + 2 \bar \mu_h^2) \bar v^2}{ \bar x^3}, \\
\left. \frac{\partial^4 V}{\partial x^2 \partial v^2}
\right|_{v=\bar v,\, x=\bar x}
&=& \frac{ 2 \mu_h^2 + \bar \mu_h^2}{ \bar x^2}, \\
\left. \frac{\partial^4 V}{\partial x^4}
\right|_{v=\bar v,\, x=\bar x}
&=& \frac{ (6 \mu_s^2 + 6 \bar \mu_s^2) \bar x^2
- (6 \mu_h^2 + 3 \bar \mu_h^2) \bar v^2}{ \bar x^4},
\eea
leading to Feynman rules depicted in Fig.~\ref{fig:feyn}. Notice that
only the cubic self-interaction of the singlet contains an odd number
of singlets such that the singlet cannot decay on tree-level.

If the singlet is much lighter than the Higgs particle, $2 m_s < m_h$,
the Higgs can decay into singlets. In this case the Higgs might even
become invisible at colliders, if the singlets have only a small decay
rate into SM particles~\cite{Bento:2000ah, Burgess:2000yq}. In the
following, we assume the singlet to be heavier than this
bound. Typically, the Higgs could also decay into two virtual singlets
and then lead to quite characteristic decays into four SM
particles. However, this is only possible if the singlet mixes with
the Higgs thus opening the singlet decay channels into SM particles
(see \cite{Barger:2007im} for a detailed discussion of this
case). Since small mixing also implies a rather long lifetime for the
singlet (see below), the most promising channel is the production of
two singlets by a Higgs with resulting missing energy signal. If the
singlets are rather heavy this requires a Higgs that is far off-shell
what makes this channel unfeasible at LHC.

\begin{figure}[ht]
\includegraphics[width=0.65\textwidth, clip ]{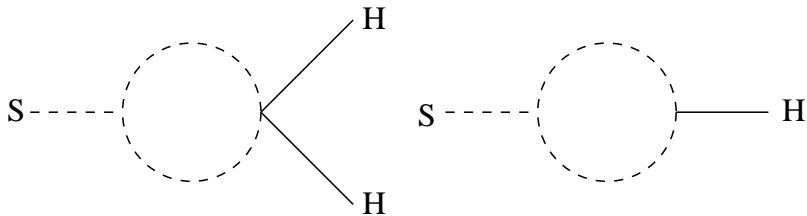}
\caption{\label{fig:decay}
One-loop diagrams that contribute to decay and mixing of the singlet
with the Higgs.}
\end{figure}

Before the discussion of the phase transition, we comment on the
possibility of the singlet being stable and hence a viable dark matter
candidate. In fact, the decay is automatically suppressed if the
mixing between Higgs and singlet is. This can be seen as follows. A
similar diagram to the decay diagram arises as a one-loop contribution
to the singlet-Higgs mixing (both depicted in
Fig.~\ref{fig:decay}). Both diagrams are in fact divergent and have to
be regularized simultaneously by adjusting the mixing parameter $\bar
\mu_{hs}^2$. This is possible since the vertex $hs$ and the vertex
$hhs$ are on tree-level as well as on one-loop proportional to each
other (with a relative factor $\bar v$ in both cases). On tree-level,
both vertices are in addition proportional to $\bar
\mu_{hs}$. This in turn implies that if the parameters are
arranged to suppress mixing, singlet decay will also be suppressed at
least by two loop orders and by the corresponding Higgs-singlet
couplings.

\begin{figure}[ht]
\includegraphics[width=0.8\textwidth, clip ]{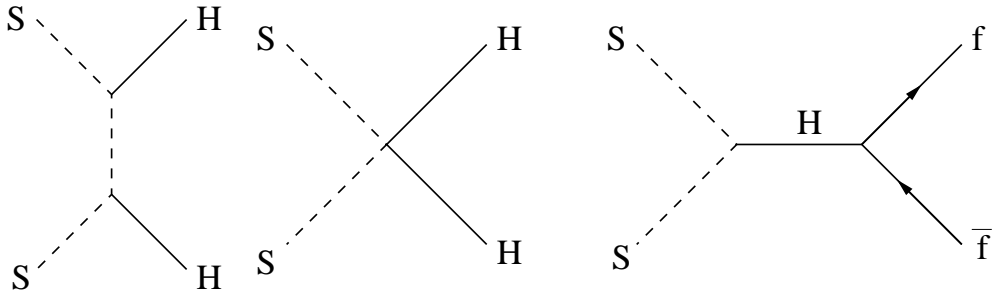}
\caption{\label{fig:ann}
Dominant contributions to singlet annihilation.}
\end{figure}

If the decay rate of the singlet vanished (or at least were smaller
than today's Hubble parameter), the annihilation rate would determine
today's singlet abundance. The annihilation cross section resulting
from the vertices $hhss$ and $hss$ are of electroweak scale which
makes the singlet a good WIMP candidate, see Fig.~\ref{fig:ann}. Due
to the WIMP miracle, the resulting singlet abundance today would then
be automatically in the right range to explain the observed dark
matter density~\cite{McDonald:1993ex}. However, in the present model,
the singlet is only protected from decaying if a (unbroken)
$\mathbb{Z}_2$ symmetry is imposed which is not the case we are
interested in. Hence, generically the singlet is not a viable dark
matter candidate.

Finally, we comment on the realization of baryogenesis in the present
model in case the electroweak phase transition is strongly
first-order. Notice that the current framework does so far not contain
a source of CP violation that is sufficient for viable electroweak
baryogenesis. The simplest way of introducing CP violation is by
resorting to dimension-six operators, as done in
ref.~\cite{Bodeker:2004ws}. Dimension-six operators that couple the
top to the Higgs field can provide sufficient CP violation without
changing collider phenomenology but give rise to electric dipole
moments in reach of the next generation of
experiments~\cite{Huber:2006ri}. Even though this seems possible, we
would like to point out that the main intent of the present work is
not to provide an example for electroweak baryogenesis with Standard
Model collider phenomenology. The aim is rather to demonstrate that
collider experiments are in general not able to determine the strength
of the phase transition. For example, a strong phase transition in the
MSSM requires some tuning in the Higgs and stop sectors such that a
similar construction as presented here also could relax these tunings
without collider traces beyond the generic MSSM (for specific collider
traces of the MSSM with a light right-handed stop see
e.g.~ref.~\cite{Menon:2009mz}). In this case the
chargino~\cite{Nelson:1991ab, Carena:2002ss, Konstandin:2005cd} or
neutralino~\cite{Cirigliano:2006dg} sector provides the required
source of CP violation.

In conclusion, the singlet model is indistinguishable from the SM by
the upcoming collider experiments as long as the mixing between the
singlet and the Higgs is small and the singlet is heavy enough to
prohibit the Higgs to decay into singlet pairs.

\section{The Phase Transition at Tree Level \label{sec_PT}}

In this section, main characteristics of the electroweak phase
transition are discussed. Most aspects of the phase transition can be
discussed while focusing on the tree-level and inclusion of one-loop
contributions (at zero temperature) lead only to minor deviations due
to the regularization scheme used. Motivated by the analysis in the
last section, we only consider the case without mixing between the
singlet and the Higgs in the broken phase, $\bar \mu^2_{hs} = 0$.

Current measurements of the electroweak precision observables favor a
rather light Higgs (at least in the case of the SM), such that
reasonable values for the Higgs mass lie in the range $\bar \mu^2_h =
(114-160 \textrm{ GeV})^2$~\cite{Barate:2003sz}. To avoid the
detection of the singlet in up-coming collider experiments, the
singlet should be heavy enough to prohibit Higgs decay into singlet
pairs, $4 \bar \mu^2_s \gtrsim \bar
\mu^2_h$. Besides it should be large enough to lead to a sizable
potential difference between the broken and the symmetric phase (to
lead to a sizable phase transition temperature) that at tree-level is
given by
\be
\Delta V = \frac{1}{12} \lp (\mu^2_h - \bar \mu^2_h ) \bar v^2 +
(\mu^2_s - \bar \mu^2_s ) \bar x^2 \rp,
\ee
what can be easily arranged by choosing the singlet mass in the broken
and symmetric phase suitably.

The singlet VEV in the broken phase, $\bar x$, constitutes an
additional free parameter. Its value is bounded from above, since a
too large path between the symmetric and the broken phase increases
the tunnel action substantially and can prohibit tunneling
altogether. On the other hand, too small values of $\bar x$ lead in
conjuncture with sizable masses to very large values of the quartic
coupling $b_4$ which would spoil the validity of perturbative
approaches. Typically, the singlet VEV is hence confined to a range
$\bar x \sim 150 - 250$ GeV.

In the present model, the strength of the phase transition is mostly
given by the tree-level parameters of the potential. This is in
contrast to e.g. the SM, where a potential barrier between the broken
and symmetric phase is only generated by thermal one-loop
contributions~\cite{Anderson:1991zb}, or to the singlet extensions
without singlet VEVs, where the barrier is generated by
Coleman-Weinberg terms~\cite{Espinosa:2007qk}. Hence, a phase
transition strong enough to avoid sphaleron
washout~\cite{Farrar:1993hn}, $v(T)/T > 1$, is easily achieved by
adjusting the tree-level parameters accordingly and is a generic
feature of the model. This is thoroughly demonstrated in
ref.~\cite{Profumo:2007wc} and can also be seen from the analysis in
the nMSSM that contains a similar scalar sector~\cite{Huber:2007vva,
Menon:2004wv}. In the following we will focus on the more exceptional
case of a phase transition that is strong enough to even give rise to
substantial gravitational radiation.

In order to analyze a first-order phase transition, we determine the
three-dimensional, Euclidean tunnel action of the bounce solution of
the scalar fields, $v(\rho)$ and $x(\rho)$,
\be
S_3 = 4 \pi \int d\rho \, \rho^2
\left[ \frac12 \lp \frac{dv}{d\rho} \rp^2
+ \frac12 \lp \frac{dx}{d\rho} \rp^2
+ V(v,x,T)
\right].
\ee
The tunneling usually proceeds for temperatures with $S_3 / T
\approx 140$ (for generalities of semi-classical tunneling see
refs.~\cite{Coleman:1977py, Callan:1977pt, Linde:1980tt}; the concrete
formulas we use in the present analysis are taken from
ref.~\cite{Huber:2007vva}).  Subsequently we determine the
characteristics of the phase transition that are relevant for
electroweak baryogenesis and the production of gravitational
radiation. These include the VEV to temperature ratio right after the
phase transition, $v / T$, that controls the suppression of sphaleron
processes, the latent heat normalized to the radiation energy of the
plasma, $\alpha$, and the inverse duration of the phase transition in
units of the Hubble parameter, $\beta/H$, that in terms of the tunnel
action are given by
\be
\alpha = \frac{30 \epsilon}{ \pi^2 g_* T^4} , \quad
\frac{\beta}{H} = T \frac{d}{dT} \frac{S_3}{T}, \nonumber 
\ee
\be
\epsilon =  V(v,x,T) - V(0,0,T) - T \frac{d }{ dT}
(V(v,x,T) - V(0,0,T)).
\ee

The tunnel action $S_3$ diverges at a temperature $T_c$ when the local
minima of the broken and symmetric phase equal in potential and is a
decreasing function in temperature. To achieve a rather strong phase
transition, it is necessary to have already at zero temperature a
potential barrier between the local minima, because otherwise the
phase transition proceeds at rather high energies, $T \approx T_c$,
and rather fast, $\beta/H \gg 100$. This potential barrier can be
achieved by choosing the corresponding parameters of the potential in
the symmetric phase positive
\be
\mu^2_s, \, \mu^2_h \,  > 0.
\ee

\begin{figure}[ht]
\includegraphics[width=0.9\textwidth, clip ]{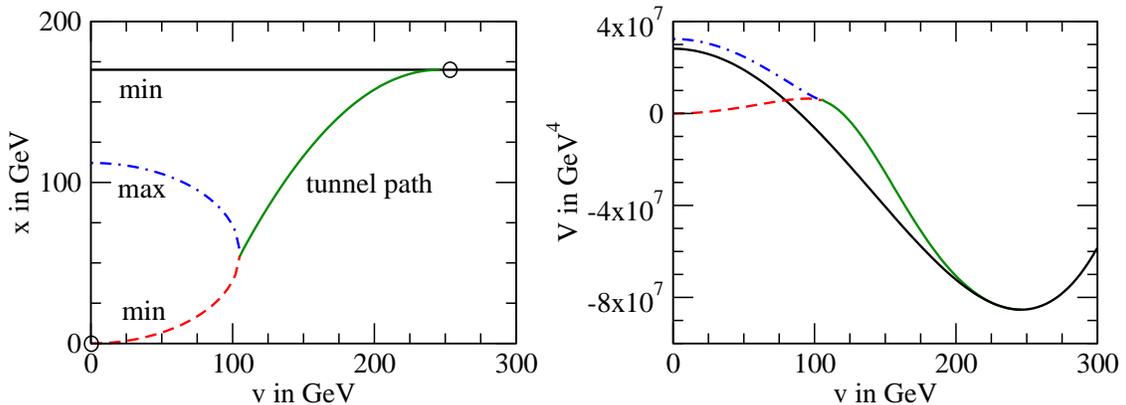}
\caption{\label{fig:path}
An example of the paths of the extrema as functions $x(v)$ in scalar
field space and the corresponding values of the potential. The two
circles denote the local minima in the potential. For small Higgs VEV
$v$, the potential has two minima and one maximum with respect to
$x$. For large $v$ only one minimum remains. The fields have to cross
the ridge to tunnel from the symmetric to the broken phase. In
addition, the plot shows a typical path we choose to determine the
tunneling action.}
\end{figure}

On the other hand, if $\mu^2_h$ is chosen rather large, the tunnel
probability is too small for the phase transition to happen, $S_3/T
\gtrsim 140$, and the model is stuck in the symmetric phase and
phenomenologically not viable. Accordingly, the Higgs mass in the
symmetric phase, $\mu^2_h$, has to be rather small for the phase
transition to happen. For one typical class of models the phase
transition proceeds as follows: At very high temperatures, the
effective potential has only one local minimum which is situated at
vanishing Higgs VEV and at a singlet VEV that ranges somewhere between
$0$ and $\bar x$. At some temperature related to the singlet mass, the
potential develops two minima $x_{1/2}$ and the minimum closer to the
origin $x_1$ is the lower of both in potential. At even lower
temperature, the minimum at $x_2$ becomes unstable and another local
minimum with non vanishing Higgs VEV develops. As soon as this new
local minimum is significantly below the one close to the origin,
tunneling might happen. This situation is depicted in
Fig.~\ref{fig:path}. In order to tunnel, the scalar fields have to
overcome a ridge between the two valleys, thus leading to a
substantial tunnel action.  A detailed numerical example including the
one-loop corrections and a numerical evaluation of the tunnel action
is given in the next section.

\section{A Numerical Example \label{sec_PTnum}}

In this section, we present the analysis of the specific numerical
example for a rather strong phase transition without traces in
upcoming collider experiments. The analysis uses the one-loop free
energy including thermal masses (to account for Daisy diagrams). The
concrete values we use are
\bea
\label{exp_params}
&\mu^2_h = (20 \, \textrm{ GeV})^2, \quad
\mu^2_s = (150 \, \textrm{ GeV})^2, \quad
\bar x = 170 \, \textrm{ GeV} , & \nonumber \\
&\bar \mu^2_s \approx (190 \, \textrm{ GeV})^2, \quad
\bar \mu^2_{hs} = (0 \, \textrm{ GeV})^2, \quad
\bar \mu^2_h = (120 \, \textrm{ GeV})^2. &
\eea

There are in principle several possibilities to determine the tunnel
action. Since, there are only two scalar fields present, one could try
to find the bounce solution of the tunnel action by combining the
methods of scanning and
over/under-shooting~\cite{Moreno:1998bq}. However, in the present case
the convergence of this algorithm is very poor, since close to the
symmetric phase the scalar fields have to follow the ridge of the
potential with high accuracy. The most sophisticated method would be
to use one of the algorithms designed to find bounce solutions in the
case of several scalar fields \cite{Konstandin:2006nd,
Cline:1999wi}. However, these methods are rather involved and an
estimate of the tunnel action is sufficient for the following
discussion. We use the following method: First we choose a path by
hand and then determine the corresponding bounce solution by
over/under-shooting. The bounce solution constitutes a saddle point of
the action with only one negative eigenvalue in the functional
determinant~\cite{Coleman:1987rm}. This negative eigenvalue is
eliminated by the over/under-shooting method while the bounce solution
is a minimum of the action with respect to changing the path in scalar
field space. One path with nearly minimal action is the one that
follows the ridge near the symmetric phase closely until its end and
then smoothly connects with the broken phase. Since the mass of the
singlet is in the present example significantly larger than the Higgs
mass, the tunnel action also prefers a path that leaves the broken
phase in the direction of the Higgs VEV. This situation is depicted in
Fig.~\ref{fig:path}.

\begin{figure}[ht]
\includegraphics[width=0.7\textwidth, clip ]{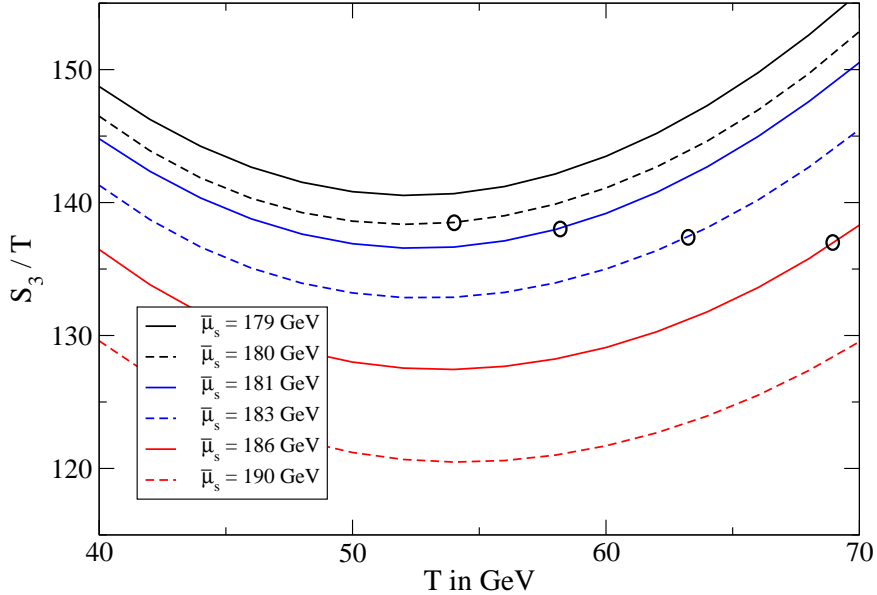}
\caption{\label{fig:S3s}
The tunnel action $S_3/T$ as a function of the temperature for
different values of $ \bar \mu_s$. The circles denote the temperature
at the end of the phase transition determined with the methods from
ref.~\cite{Huber:2007vva}. }
\end{figure}

In Fig.~\ref{fig:S3s}, the tunnel action is depicted as a function of
temperature for several values of $\bar \mu^2_s$ and the remaining
parameters as in eq.~(\ref{exp_params}). By adjusting $\bar \mu^2_s$,
a model can be found in which the scalar fields barely tunnel. The
phase transition proceeds in this case very slowly and with only a few
bubbles per Hubble volume and a rather large latent heat results. The
corresponding characteristics of the phase transition are given in
table~\ref{tab_chara}.

\begin{table}
\begin{tabular}[b]{|c||c|c|c|c|}
\hline
\quad $\bar \mu_s/$ GeV \quad&
\quad $\alpha$ \quad\quad &
\,\, $\beta/H$ \quad &
\,\, $v/T$ \quad &
\,\, $T$ / GeV \quad  \\
\hline
\hline
190 & 0.14 & 121 & 3.1 & 75 \\
\hline
186 & 0.18 & 88 &  3.4 & 69 \\
\hline
183 & 0.25 & 53 & 3.7 & 63 \\
\hline
181 & 0.33  & 25 & 4.0 & 57 \\
\hline
180 & 0.42  & 8 & 4.2 & 54 \\
\hline
179 &
\multicolumn{4}{|c|}
{symmetric phase stable}
\\
\hline
\end{tabular}
\caption{Sets of parameters corresponding to Fig.~\ref{fig:S3s}.
\label{tab_chara}
}
\end{table}

\begin{figure}[ht]
\includegraphics[width=0.6\textwidth, clip ]{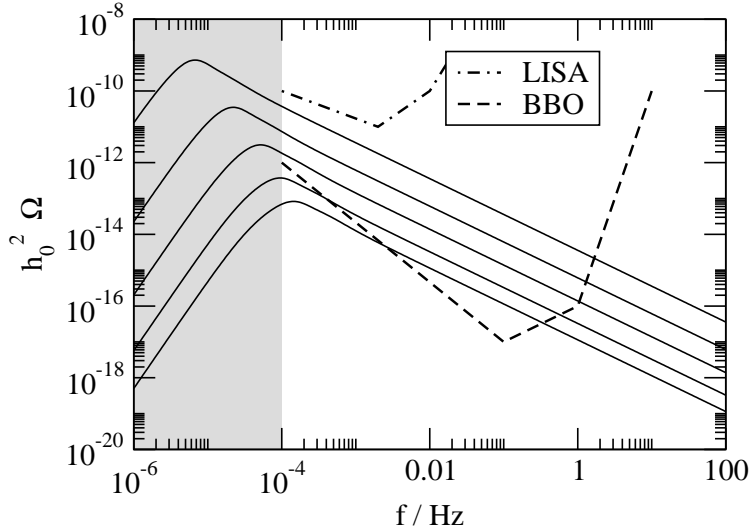}
\caption{\label{fig:GWs}
The spectra of gravitational wave generation for the parameters given
in Tab.~\ref{tab_chara} (the parameter $\alpha$ is increasing from
bottom to top). The dashed lines show the sensitivity of the LISA and
BBO experiments. In the shaded region the sensitivity of both
experiments drops significantly.}
\end{figure}

In all cases the phase transition is strong enough for viable
baryogenesis, $v/T \gtrsim 1$. Finally we would like to comment on
gravitational wave production during the phase transition. For a
general discussion of the different arising contributions to
gravitational wave production we refer the reader to
refs.~\cite{Nicolis:2003tg, Grojean:2006bp, Huber:2007vva}. For
gravitational wave production observable by the the planned Big Bang
Observer~\cite{Corbin:2005ny}, a phase transition at electroweak
scales typically requires $\alpha
\gtrsim 0.1$ and $\beta/H \lesssim 200$ for a signal from bubble
collisions~\cite{Kosowsky:1992vn, Huber:2008hg} and a slightly
stronger phase transition for a signal from turbulence
\cite{Dolgov:2002ra, Caprini:2006jb}. The gravitational wave spectra
resulting from colliding bubbles as given in ref.~\cite{Huber:2008hg}
and from turbulence as given in refs.~\cite{Huber:2007vva,
Caprini:2006jb} and for the parameters in Tab.~\ref{tab_chara} are
shown in Fig.~\ref{fig:GWs}.

We conclude that a phase transition strong enough for electroweak
baryogenesis is quite generic in the present model. Additionally, with
a tuning in the parameter $\bar \mu_s$ on the few percent level, a
model can be found that even can lead to gravitational wave production
observable by the Big Bang Observer.

\section{Discussion \label{sec_concl}}

Main concern of the present work is the question if experimental
insights about the Higgs sector from upcoming collider experiments
will be sufficient to reliably predict the nature of the electroweak
phase transition. In order to do so we discussed a singlet extension
of the SM as a showcase. Our main conclusion is that an extremely
strong first-order phase transition could have occurred in the early
Universe without leading to any collider signals beyond the SM.

Main constraint on the model is that the mixing between the Higgs and
the singlet is suppressed. Besides, the singlet has to be heavy enough
to avoid Higgs decay into singlet pairs. Interestingly, the singlet is
under these assumptions (rather) stable which further reduces the
chances of the model to be distinguished from the SM at colliders. On
the other hand, vanishing mixing between the Higgs and the singlet is
not protected by any symmetry in the present model (and also not
stable under radiative corrections) such that mixing is not expected
to vanish exactly and should be experimentally testable at some
stage. The prospects to distinguish the present model from the SM are
hereby significantly higher with the next generation of linear
collider experiments than with the LHC if mixing is small.

With the tree-level parametrization we used in the present analysis,
regions in parameter space with a first-order phase transition are
easily found:

\begin{itemize}
\item
A first-order phase transition strong enough for electroweak
baryogenesis, $\phi/T \gtrsim 1$, is a quite generic feature of the
present model. 

\item
A first-order phase transition strong enough for a gravitational wave
production that is at least observable by the Big Bang Observer
typically requires a rather small Higgs mass in the symmetric phase
and a tuning of the singlet masses (in the symmetric or broken
phases) on the few percent level.
\end{itemize}

We would like to emphasize that in the present analysis we focused
from the beginning on regions in parameter space with no collider
signals beyond the SM. Nevertheless, in the general case it is found
that regions in parameter space with strong phase transitions can be
correlated with specific collider signatures as e.g. exotic final
states~\cite{Profumo:2007wc, Barger:2007im}.

Finally, notice that the model discussed here is most probably the
minimal model that reconciles SM collider phenomenology with a strong
first-order phase transition. The essential ingredient is hereby that
the new scalar field has a changing VEV during the electroweak phase
transition. If fields are added to the SM that are coupled strongly to
the Higgs but do not obtain a VEV, the Higgs self-couplings are
typically modified compared to the SM value by the same one-loop
contributions to the scalar potential that strengthen the phase
transition. This is basically the argument put forward in
ref.~\cite{Noble:2007kk} supporting the claim that a strong phase
transition will generically leave significant traces in the
interactions of the Higgs sector. The present work provides a concrete
counterexample to this claim.

\section*{Acknowledgments}
We would like to thank J.~R.~Espinosa and A.~T.~Pierce for many
helpful discussions.  T.~K. acknowledges support by the Marie Curie
Research \& Training Network 'UniverseNet'
(MRTN-CT-2006-035863). A.~A. is supported by NSERC of Canada and MCTP.

\appendix

\section{Reparametrization}

In this section, we display the specific reparametrization of the
parameters in the potential. In order to use this description also for
the determination of the counterterms, we define in the broken phase
the first derivatives
\be
\label{constr1A}
\bar d_h = \left. \frac{\partial V}{\partial v}
\right|_{v=\bar v,\, x=\bar x}, \quad
\bar d_s = \left. \frac{\partial V}{\partial x}
\right|_{v=\bar v,\, x=\bar x},
\ee
the tree-level Higgs and singlet masses and mixing
\be
\label{constr2A}
\bar \mu_h^2 = \left. \frac{\partial^2 V}{\partial v^2} \right|_{v=\bar v,\, x=\bar x}, \quad
\bar \mu_s^2 = \left. \frac{\partial^2 V}{\partial x^2} \right|_{v=\bar v,\, x=\bar x}, \quad
\bar \mu_{hs}^2 = \left. \frac{\partial^2 V}{\partial v \partial x} \right|_{v=\bar v,\, x=\bar x},
\ee
the first derivative in the symmetric phase and the second derivatives
of the potential in the symmetric phase
\label{constr3A}
\be
d_s = \left. \frac{\partial V}{\partial x} \right|_{v=0,\, x=0}, \quad
\mu_h^2 = \left. \frac{\partial^2 V}{\partial v^2} \right|_{v=0,\, x=0}, \quad
\mu_s^2 = \left. \frac{\partial^2 V}{\partial x^2} \right|_{v=0,\, x=0}.
\ee
The resulting parameters then read
\bea
b_1 &=& d_s,\quad
\mu^2 = -\mu^2_h,\quad
b_2 = \mu^2_s, \quad
\lambda_0 = -\frac{\bar d_h - \bar \mu^2_h \bar v}{2 \bar v^3}, \\
a_1 &=& -\frac{-6 \bar d_h + 4 \mu^2_h \bar v +
2 \bar \mu^2_h \bar v + 2 \bar \mu^2_{hs} \bar x}{\bar v \bar x}, \quad
a_2 = \frac{-3 \bar d_h + 2 \mu^2_h \bar v
+ \bar \mu^2_h \bar v + 2 \bar \mu^2_{hs} \bar x}{\bar v \bar x^2}, \\
b_3 &=& -\frac{ 3 \bar d_h \bar v - 2 \mu^2_h \bar v^2
- \bar \mu^2_h \bar v^2 + 6 d_s \bar x -
6 \bar d_s \bar x + \mu^2_{hs} \bar v \bar x
+ 4 \mu^2_s \bar x^2 + 2 \bar \mu^2_s \bar x^2}{
   2 \bar x^3}, \\
b_4 &=& -\frac{-3 \bar d_h \bar v + 2 \mu^2_h \bar v^2
+ \bar \mu^2_h \bar v^2 - 4 d_s \bar x
+ 4 \bar d_s \bar x - 2 \mu^2_s  \bar x^2 - 2 \bar \mu^2_s \bar x^2}
{2 \bar x^4}.
\eea

From these formulas it is clear that the limit $\bar x \to 0$ leads in
certain cases to a divergent set of parameters. The very same
relations can be used to determine the counterterms as functions of
the derivatives of the one-loop potential.

\section{Thermal Masses}

The scalar masses are derived from the scalar potential as given in
eq.~(\ref{pot_old})
\be
V = - \frac{\mu^2}{2} \, v^2 + \frac{\lambda_0}{4} v^4
+ \frac{a_1}{4} \, v^2 \, x + \frac{a_2}{4} v^2 \, x^2
+ b_1 x + \frac{b_2}{2} x^2 + \frac{b_3}{3} x^3 + \frac{b_4}{3} x^4.
\ee
and can be recast using the relations in the previous section. The
Higgs/singlet mass matrix reads
\be
m^2_{HS} =
\begin{pmatrix}
 - \mu^2 + 3 \lambda_0 v^2 + \frac12 a_1  x + \frac12 a_2 x^2 &
\frac12 a_1  v + a_2 \, v \, x \\
\frac12 a_1 v + a_2 \, v \, x &
\frac12 a_2  v^2 + b_2 + 2 b_3\,x + 3 b_4\,x^4\\
\end{pmatrix}
\ee
and the Goldstone masses are given by
\be
m^2_G =  - \mu^2 + 3 \lambda_0 v^2 + \frac12 a_1  x + \frac12 a_2 x^2.
\ee
The masses of the gauge bosons are
\be
m^2_{gb} =
\begin{pmatrix}
\frac{g^2}{4} v^2 & 0 & 0 & 0 \\
0 & \frac{g^2}{4} v^2 & 0 & 0 \\
0 & 0 & \frac{g^2}{4} v^2 & \frac{g g^\prime}{4} v^2 \\
0 & 0 & \frac{g g^\prime}{4} v^2 & \frac{g^{\prime 2}}{4} v^2 \\
\end{pmatrix}
\ee
and the top quark mass is given by
\be
m_t^2 = \frac{y_t^2}{2} v^2.
\ee
We take account of the corresponding self-energies at finite
temperature in leading order. For the Higgs/singlet bosons they read
\be
\Pi_{HS} =
\begin{pmatrix}
\left(\frac{3}{16} g^2 + \frac{1}{16} g^{\prime 2} + \frac12 \lambda_0
+ \frac14 y_t^2 + \frac{1}{24} a_2 \right) T^2 & 0 \\
0 & \left(\frac14 b_4 + \frac{1}{6} a_2 \right) T^2 \\
\end{pmatrix},
\ee
and for the Goldstone bosons
\be
\Pi_G = \left(\frac{3}{16} g^2 + \frac{1}{16} g^{\prime 2} + \frac12 \lambda_0
+ \frac14 y_t^2 + \frac{1}{24} a_2 \right) T^2.
\ee
Finally, for the longitudinal gauge bosons one finds
\be
\Pi_{gb} =
\begin{pmatrix}
\frac{11}{6} g^2 T^2 & 0 & 0 & 0 \\
0 & \frac{11}{6} g^2 T^2 &  0 & 0 \\
0 & 0 & \frac{11}{6} g^2 T^2 & 0 \\
0 & 0 & 0 & \frac{11}{6} g^{\prime 2} T^2 \\
\end{pmatrix}.
\ee

\end{document}